\documentclass[11pt]{article}

\usepackage{geometry}  
\usepackage[T1]{fontenc}  
\usepackage[utf8x]{inputenc}  
\usepackage{lineno}  
\usepackage{fancyhdr}  
\usepackage{enumitem}  
\usepackage{hyperref}  
\usepackage{titling}  
\usepackage{natbib}  
\usepackage{mathtools}  
\usepackage{titlesec}  
\usepackage{lastpage}  
\usepackage{tabularx}
\usepackage{booktabs}

\usepackage[english]{babel}  
\usepackage{amsmath}  
\usepackage{wasysym}  
\usepackage{bbm}  
\usepackage{array}  
\usepackage{xr}  
\usepackage{verbatim} 
\usepackage{float} 
\usepackage{xcolor}
\usepackage{multirow}

\usepackage{amsmath,amssymb,amsfonts}
\usepackage{algorithm}
\usepackage{algpseudocode}
\usepackage{graphicx}
\usepackage{textcomp}

\def\cN{{\cal N}}
\def\R{{\mathbb R}}
\def\C{{\mathbb C}}

\DeclareMathOperator*{\argmin}{argmin\,}
\newcommand\norm[1]{\lVert#1\rVert}
\newcommand\nth[1]{$#1^{\mathrm{th}}$}
\renewcommand\bold[1]{\mathbf{#1}}

\definecolor{blue1}{RGB}{79, 113, 190}
\definecolor{orange1}{RGB}{239, 137, 51}
\definecolor{graynew}{gray}{0.88}
\definecolor{purple1}{RGB}{112, 48, 160}

\hypersetup{%
  pdfborderstyle={/S/U/W 1}
}

\setcitestyle{round,numbers}
\titleformat{name=\section}{\normalfont\Large\bfseries}{}{0pt}{}
\titleformat{name=\subsection}{\normalfont\large\bfseries}{}{0pt}{}
\titleformat{name=\subsubsection}{\normalfont\normalsize\bfseries}{}{0pt}{}

\newcommand{\email}[1]{\href{mailto:{#1}}{{#1}}}

\newcommand{\keywords}[1]{\textbf{Keywords}: {#1}}

\newcommand{\optincludegraphics}[2][]{}

\newcommand{\optinput}[1]{}

\newtagform{brackets}{[}{]}
\usetagform{brackets}

\newcommand{\thejournal}[1]{Magnetic Resonance in Medicine}

\title{Accelerated Motion Correction with Deep Generative Diffusion Models}

\immediate\write18{texcount -sum=1,1,1,1,1,1,1 -merge -incbib -dir -utf8 \jobname.tex > \jobname.wcTotal }
\immediate\write18{texcount -sub=none -merge -incbib -dir -utf8 \jobname.tex | grep _main_ | grep -oE '[0-9]+' | python -c "import sys; print(sum(int(l) * w for l, w in zip(sys.stdin, [1, 1, 0, 0, 0, 0, 0])))" > \jobname.wcManuscript }
\immediate\write18{texcount -sub=none -merge -incbib -dir -utf8 \jobname.tex | grep Abstract | grep -oE '[0-9]+' | python -c "import sys; print(sum(int(l) * w for l, w in zip(sys.stdin, [1, 1, 0, 0, 0, 0, 0])))" > \jobname.wcAbstract }

\newcommand{\wcTotal}{\clearpage{\noindent\large{\bf Detailed Word Count} (not to be included for submission)}\verbatiminput{\jobname.wcTotal}}
\newcommand{\wcManuscript}{\input{\jobname.wcManuscript}}
\newcommand{\wcAbstract}{\input{\jobname.wcAbstract}}

\begin{document}

\begin{titlepage}
{\noindent\LARGE\bf \thetitle}

\bigskip

\begin{flushleft}\large
	Brett Levac\textsuperscript{1,*},
	Sidharth Kumar \textsuperscript{1},
	Ajil Jalal\textsuperscript{2},
	Jonathan I. Tamir\textsuperscript{1},

\end{flushleft}

\bigskip

\noindent
\begin{enumerate}[label=\textbf{\arabic*}]
\item{Chandra Family Department of Electrical and Computer Engineering, The University of Texas at Austin, Austin, TX, USA}
\item {Electrical Engineering and Computer Sciences, University of California at Berkeley, California, USA}
\end{enumerate}

\bigskip


\textbf{*} Corresponding author:

\indent\indent
\begin{tabular}{>{\bfseries}rl}
Name		& Brett Levac 
\\
Department	& Chandra Family Department of Electrical and Computer Engineering													\\
Institute	& University of Texas at Austin														\\
         	& Texas														\\
			& 78712
			\\
            & United States														\\
E-mail		& \email{blevac@utexas.edu}											\\
\end{tabular}

\vfill

Approximate word count: 120 (Abstract) 4200 (body)

Portions of this work were presented at IEEE International Symposium on Biomedical Imaging (ISBI) 2023 \cite{levac_isbi}, and ISMRM Workshop on Data Sampling and Image Reconstruction 2023 \cite{levac_sedona}.
\end{titlepage}

\pagebreak

\begin{abstract}


\noindent\textbf{Purpose:} The aim of this work was to develop a method to solve the ill-posed inverse problem of accelerated image reconstruction while correcting forward model imperfections in the context of subject motion during MRI examinations.

\noindent\textbf{Methods:} 
The proposed solution uses a Bayesian framework based on deep generative diffusion models to jointly estimate a motion-free image and rigid motion estimates from sub-sampled and motion-corrupt k-space data.

\noindent\textbf{Results:} We are able to reconstruct motion-free images from accelerated 2D Cartesian and non-Cartesian scans without any external reference signal. We show that our method improves over existing  correction techniques on both simulated and prospectively accelerated data.

\noindent\textbf{Conclusion:} 
We demonstrate a flexible framework for retrospective motion correction of accelerated MRI based on deep generative diffusion models.
 
\noindent\keywords{MRI, Accelerated Reconstruction, Motion Correction, Deep Generative Diffusion Model}
\end{abstract}

\section{Introduction}
\label{sec:intro}
MRI is a highly effective medical imaging modality which owes much of its utility to having superior soft tissue contrast without any ionizing radiation. Unfortunately, MRI is notoriously slow when compared to other imaging methods. This limitation leads to increased operating costs and even decreased image quality due to a variety of factors. A common way to reduce scan time is to simply acquire less data and thus subsample k-space. This process, however, makes the task of recovering the desired image an ill-posed inverse problem. To better handle this task, many techniques have been developed such as parallel imaging \cite{sense,smash,grappa}, handcrafted image regularization \cite{lustig2007sparse,LORAKS,trzasko}, dictionary learning \cite{single_dict}, subspace constraints \cite{T2Shuff_MRM}, and more recently deep learning \cite{hammernik, modl,deepjsense, CSGM-bora17,jalal2021robust, martin_score, CHUNG_score}.

\begin{figure}[h!]
\begin{center}
\includegraphics[width=1.\linewidth]{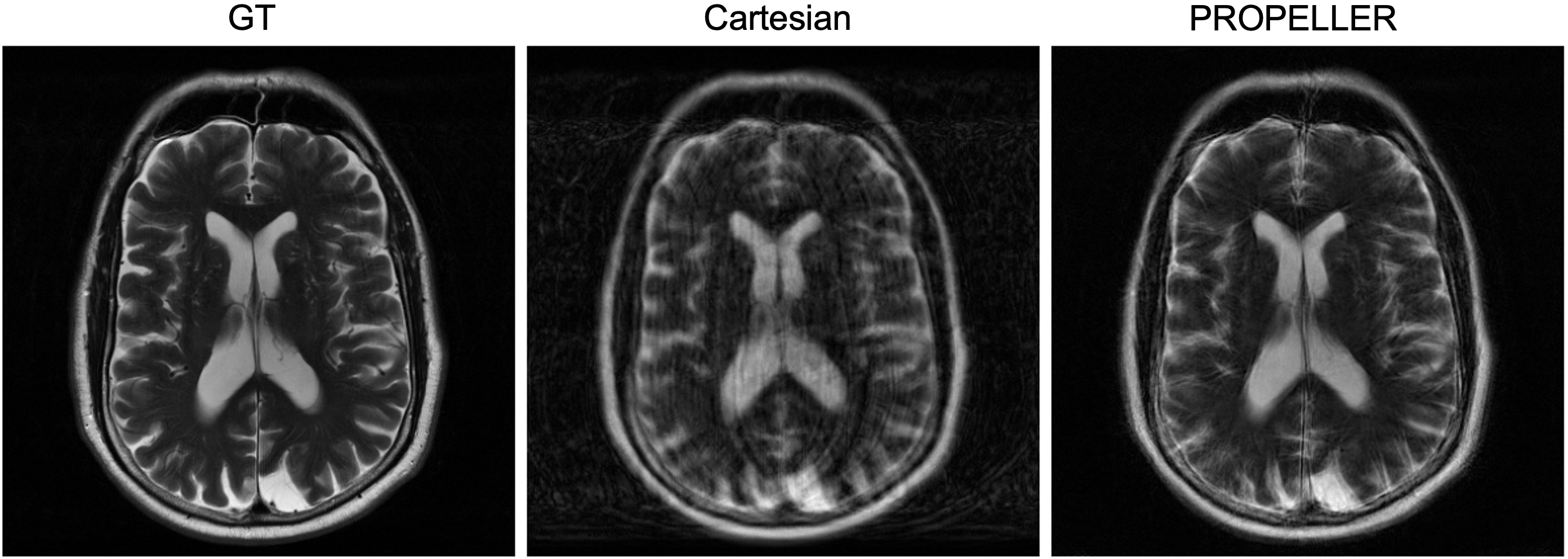}
\end{center}
\caption{(Left) Motion-free image, (Center) simulated motion-corrupt image resulting from Cartesian fast spin-echo ($R=1$), (Right) simulated motion-corrupt image resulting from PROPELLER fast spin-echo ($R=1$). Both motion-corrupt images were fully sampled and the same motion states were used for each TR.}
\label{fig:sampling_effects}
\end{figure}

Although highly subsampling measurements reduces the likelihood of motion occurring during the scan, MRI is still susceptible to subject motion due to physical constraints during a given scan such as the repetition time (TR) needed between excitations. The resulting artifacts can often render the image non-diagnostic, and may ultimately require the corrupted scan to be reqacquired \cite{zaitsev2015motion}. The severity of motion artifacts in the final image is related to a variety of factors such as acquisition parameters (sampling trajectory, echo train ordering, signal preparation) and the degree of motion. See Fig. \ref{fig:sampling_effects} for an example. These artifacts have tangible costs in clinical settings, especially when scanning pediatric patients where motion artifacts are extremely common \cite{slipsager2020}. 
Many approaches to address motion corruption have been proposed. These methods can be separated into two broad categories: prospective, and retrospective. 

Prospective methods are categorized as those which can be used at scan time to modify the acquisition in response to patient motion. To measure motion during the scan a variety of approaches have been proposed that either leverage additional pulse sequence actions like motion navigators or external measurement devices such as respiratory bellows, PILOT tones, nuclear magnetic resonance probes, and optical tracking \cite{Butterfly,Pruessmann,PILOT,nael2021prospective,prospective_survey}. These measurements can be used to correct motion by binning data into different motion states to later inform the reconstruction process, or even discarding corrupted measurements and guiding reacquisition of corrupted data. Due to the overhead created by additional measurement equipment and reacquisition of data, these methods may still increase operating costs and even scan time, as well as require modifications to the sequence or addition of peripheral hardware.

Retrospective methods assume no control of the imaging procedure and correct for motion artifacts after measurement data have been collected. This means  techniques that require sequence modification are not possible. They also typically assume no access to direct measurements of the true motion states which occurred during the exam. Retrospective techniques are more widely applicable but also face a more difficult task than prospective methods.

In light of recent advancements in the area of deep learning, perhaps the most straightforward approach to retrospectively correct motion is to directly map motion corrupt images to clean images. One such approach successfully trained a conditional generative adversarial network (GAN) to translate motion corrupt images to clean images \cite{johnson2019}. This technique falls into the class of end-to-end deep learning methods. However, as previously stated, artifact appearance is heavily dependent on the chosen forward operator, i.e. the sampling trajectory (Fig. \ref{fig:sampling_effects}).  Due to this, a network's performance at test time is highly dependent on how motion is synthesized at training time to create training pairs \cite{FSE_comp_moco}. 

Motion can be described as an unknown perturbation to the assumed forward model that gives rise to artifacts at reconstruction time. This has led previous works to jointly solve an optimization problem for the target image and the unknown motion that occurred at scan time \cite{sense_encode,lingala,Rizzuti,DISORDER,tamer}. These methods have primarily been applied to the low acceleration regime. To build upon joint optimization, supervised learning has been used as one step in a larger iterative algorithm that jointly solves for the image and the motion parameters \cite{namer}. Although this method shows notable improvements over prior methods, it is still likely susceptible to distribution shifts in the forward operator (changes to acquisition and sampling parameters), as to train the end-to-end network component it is necessary to pre-select the manifestation of the motion artifacts to learn the proper inversion. It also still relies on a linear reconstruction backbone for solving the accelerated reconstruction task which is not as powerful as recent deep learning based reconstruction techniques.

In contrast to these techniques, in this work we propose a retrospective motion correction technique that builds off of recent advances in deep generative models \cite{song_ermon,DDPM}. In particular, we formulate the reconstruction under the lens of posterior sampling \cite{jalal2021robust,CHUNG_score,martin_score}. We extend the framework to joint posterior sampling over the image and the rigid motion parameters.
Our goal was to develop a method which is \textbf{(1)} effective at correcting in-plane, rigid motion from subsampled data while \textbf{(2)} being agnostic to choices in the forward model which can greatly affect the manifestation of the motion artifacts observed.

\section{Theory}

\subsection{Accelerated MRI Reconstruction}
The goal of accelerated image reconstruction in MRI is to recover an image $\mathbf{x}\in \C^N$ from under-sampled Fourier measurements (k-space) $\mathbf{y}\in \C^M$. We can denote the measurement (forward) process in MRI as
\begin{equation}
\label{eqn:simple_forw}
\mathbf{y}^{(l)} = N_K S_l \mathbf{x} + \eta_l,
\end{equation}
where $\mathbf y^{(l)}$ is k-space of the \nth{l} coil, $N_K \in \C^{M\times N}$ denotes the non-uniform Fourier transform operator (2D or 3D) evaluated at coordinates $K\in \R^{d\times M}$, $d=2,3$ for 2D and 3D imaging respectively, $S_l \in \C^{N \times N}$ is the \nth{l} coil sensitivity map, and $\eta_l \sim \cN(0,\sigma^2 I)$ is additive noise. We can consolidate the forward operator for all coils into one operator $A = N_{K}S$.

Viewed from the perspective of regularized inverse problems, reconstruction can then be formulated as solving the optimization problem
\begin{equation}
\label{eqn:recon_opt}
\mathbf{x}^* = \argmin_{\mathbf{x}} \norm{A\mathbf{x}-\mathbf{y}}_2^2 + \lambda R(\mathbf{x}),
\end{equation}
where $R(\mathbf{x})$ can be a handcrafted image regularization term such as L1-wavelet sparsity \cite{lustig2007sparse}, or low-rank structure \cite{LORAKS,trzasko}. Reconstruction can also be solved with deep networks by learning a mapping ($f_\theta$) from measurement to image space using training data\cite{modl,hammernik}
\begin{equation}
\mathbf{x}^* = f_\theta(\mathbf{y}).
\end{equation}

More recently, there has been a push to use deep generative models to learn useful statistical priors for regularization \cite{tezcan_bayesianmri,jason_bayesianmri,jalal2021robust,CHUNG_score,martin_score}. In these techniques, reconstruction takes on a Bayesian formulation where the goal is to solve the inverse problem with a variety of estimators such as maximum a posteriori (MAP) estimation, minimum mean square error (MMSE) estimation, or posterior sampling which have perceptual benefits over many other formulations \cite{Blau_2018_CVPR}. 

\subsection{Diffusion Generative Models for Inverse Problems}
Recent work in the generative model space has been focused on diffusion processes \cite{song_ermon,DDPM,song2021sde,song2021sde,karras2022edm}.  For the remainder of this section we will adopt the notation introduced in \cite{karras2022edm}. Diffusion generative models can be understood through viewing two complementary stochastic differential equations (SDE). The first SDE is called the forward process. In the forward process, noise is gradually added to the data distribution of interest:
\begin{equation}
    \label{eqn:sde_forward}
    d\bold{x} = \frac{s'(t)}{s(t)}\bold{x}dt + s(t)\sqrt{2\sigma^{'}(t)\sigma(t)}d\bold{\omega}.
\end{equation}
Here $\frac{s'(t)}{s(t)}$ is often referred to as the drift coefficient, while $s(t)\sqrt{2\sigma^{'}(t)\sigma(t)}$ is commonly called the diffusion coefficient, and $\bold{\omega}$ defines a Brownian motion process. This process can be reversed via a complementary SDE or ordinary differential equation (ODE) \cite{Anderson1982ReversetimeDE,song2021sde,karras2022edm}. We will focus on the reverse ODE which is given by:
\begin{equation}
    \label{eqn:ode_reverse}
    d\bold{x} = \left(\frac{s'(t)}{s(t)}\mathbf{x}-s^2(t)\sigma'(t)\sigma(t)\nabla_\bold{x}\log {p \left(\Tilde{\mathbf{x}};\sigma(t)\right)}\right)dt
\end{equation}
where $\Tilde{\mathbf{x}}=\frac{\mathbf{x}}{s(t)}$. When run from time $t=T$ to $t=0$ the procedure results in sampling from the original clean data distribution $p_{data}(\bold{x})$. Here we note that $s(t)$ and $\sigma(t)$ are analytically defined in the forward equation (Eq. \ref{eqn:sde_forward}), so the only portion of the reverse ODE which needs to be learned is the score ($\nabla_\bold{x}\log{p(\bold{x};\sigma(t))}$) at each time $t$. The score can be approximated by training a neural network ($D_\theta(\mathbf{x}, \sigma(t))$) via denoising score matching \cite{denoise_score}. For clarity in future equations, we note here that $D_\theta(\bold{x};\sigma(t))$ is not a direct approximation to the score function but rather is trained to predict the denoised signal at each noise level leading to the following relation with the score function at each time point during the reverse process:
\begin{equation}
\nabla_{\bold{x}} \log{p(\bold{x};\sigma(t))} = \frac{ D(\bold{x};\sigma(t))- \bold{x}}{\sigma^2(t)}.
\end{equation}

With access to an approximation of the true score function, the reverse ODE can be solved using ODE solvers like Euler's method (1st order). To solve inverse problems, we can instead use the following reverse ODE:
\begin{equation}
    \label{eqn:posterior_ode_reverse_bayes}
    d\bold{x} = \left(\frac{s'(t)}{s(t)}\bold{x} \right. - s^2(t)\sigma'(t)\sigma(t)\left[ \nabla_\bold{x}\log{p(\bold{y}|\Tilde{\bold{x}};\sigma(t))} \right.  + \left. \nabla_\bold{x}\log{p(\Tilde{\bold{x}};\sigma(t))}\right] \left. \vphantom{\frac{s'(t)}{s(t)}} \right)dt
\end{equation}
Following this ODE, we will be sampling from the posterior distribution $p(\bold{x}|\bold{y})$. The key issue with this approach is that we only analytically know the form of the likelihood at time $t=0$ (e.g. $p(\mathbf{y}|\mathbf{x};\sigma(0))=\mathcal{N}(A\mathbf{x},\sigma^2 I)$). Prior works like Diffusion Posterior Sampling (DPS) \cite{chung2023dps} have approximated the likelihood at intermediate times steps with 
\begin{equation}
    \label{eqn:DPS_approx_liklihood}
    p(\mathbf{y}|\mathbf{x}(t);\sigma(t))\approx p(\mathbf{y}|\hat{\mathbf{x}};\sigma(0)), 
\end{equation}
where $\hat{\bold{x}} = E[\mathbf{x}(0)|\mathbf{x}(t)]$ is an estimate of the denoised image at time $t=0$ and is given by Tweedie's formula \cite{tweedie} to be 
\begin{align}
     \label{eqn:DPS_edm_expecation}
    E[\bold{x}(0)|\bold{x}(t)] = \frac{1}{s(t)}\left(\bold{x}(t) + s^2(t)\sigma^2(t) \nabla_{\bold{x}} \log{p(\bold{x}(t);\sigma(t))}\right)\\
    = \frac{1}{s(t)}\left(\bold{x}(t) + s^2(t) \left(D_\theta(\bold{x}(t);\sigma(t)) - \bold{x}(t)\right) \right)
\end{align}
This is leads to an inference procedure for solving inverse problems as shown in Alg. \ref{alg:DPS}.


\begin{algorithm}[t]
\caption{Diffusion Posterior Sampling \cite{chung2023dps}}\label{alg:DPS}
\begin{algorithmic}[1]
\Procedure{PS}{$D_\theta(\bold{x};\sigma(t)),\sigma(t),s(t), t_{i\in{0,...,N}}, \mathbf{y}$}
\State \textbf{sample}
$\mathbf{x}(t_0) \sim \cN(\mathbf{0}, \sigma^2(t_0)s^2(t_0)\mathbf{I})$
  \For{$i \in \{0,...,N-1\}$}
  \State $\Tilde{\mathbf{x}} = \frac{\mathbf{x}(t_i)}{s(t_i)}$
  
  \State $\nu_i = (\frac{\sigma'(t_i)}{\sigma(t_i)}  + \frac{s'(t_i)}{s(t_i)})$
  \State $\zeta_i = \frac{\sigma'(t_i)s(t_i)}{\sigma(t_i)}$

  \State $\mathbf{d}_P = \nu_i\mathbf{x}(t_i) -  \zeta_{i} D_\theta\left(\Tilde{\mathbf{x}};\sigma(t_i) \right)$

  \State $\hat{\mathbf{x}} = \frac{1}{s(t_i)}\left(\mathbf{x}(t_i) + s(t_i)^2 \sigma(t_i)^2 \frac{D_\theta\left(\Tilde{\mathbf{x}};\sigma(t_i)\right)-\mathbf{x}(t_i)}{\sigma(t_i)^2}\right)$

  \State $\mathbf{d}_L = \nabla_{\mathbf{x}} \norm{A\hat{\mathbf{x}}(\mathbf{x}(t_i))-\mathbf{y}}_2^2$

  \State $\mathbf{x}(t_{i+1}) \leftarrow \mathbf{x}(t_i) + (t_{i+1}-t_i)\bold{d}_P + \gamma_{t_i} \mathbf{d}_L$
  \EndFor
\EndProcedure
\end{algorithmic}
\end{algorithm}

\subsection{Measurement Formation in the Presence of Motion}
\label{subsec:motion_sim}
We consider motion which is rigid, in-plane, and occurs between readout lines. The assumption that motion does not occur during the readout period is not too restrictive since the readout duration is typically much shorter than the time between readouts. This assumption means that issues such as spin-history effects are not considered. Under these assumptions we can characterize the effects of rigid body motion (rotation, translation) on k-space measurements using simple Fourier theory. In particular, rotation in image space leads to the same rotation in k-space, while translation in image space causes linear phase shifts in k-space. Both of these effects can be captured in a modified forward operator:
\begin{equation}
\label{eqn:motion_forw_expanded}
    \mathbf{y}_i = P_{\phi_i}N_{R_{\theta_i}K_i}S\mathbf{x} + \eta, \text{ }\eta \sim \mathcal{N}(0, \sigma^2 I),
\end{equation}
where $\mathbf{x} \in \mathbb{C}^N$ is the motion-free image, $S \in \mathbb{C}^{N_cN\times N_cN}$ contains the $N_c$ sensitivity maps, $R_{\theta_i}$ is a rotation matrix for the \nth{i} motion state, $P_{\phi_i}$ is a diagonal matrix implementing a linear phase shift describing the horizontal and vertical translations during the \nth{i} motion state, $K_i$  are the coordinates for the intended k-space trajectory during the \nth{i} motion state, and $N_{R_{\theta_i}K_i}$ is the Non-uniform  Fast Fourier Transform (NUFFT) of $S\mathbf{x}$ at the coordinates $R_{\theta_i}K_i$. We note here that although $\mathbf{y}_i$ and $\mathbf{x}$ are linearly related, $\phi_i$, $\theta_i$ and $\mathbf{y}_i$ are not.

For ease of notation we combine all motion states for the rotation angles, translation distances and intended sampling trajectories into the variables $\theta$, $\phi$, and $K$ respectively:
\begin{equation}
\label{eqn:motion_forw_less_expanded}
    \mathbf{y} = P_{\phi}N_{R_{\theta}K}S\mathbf{x} + \eta, \text{ }\eta \sim \mathcal{N}(0, \sigma^2 I).
\end{equation}
To further simplify this expression we combine all unknown motion parameters ($\theta$, $\phi$) into $\kappa$ and get the following expression
\begin{equation}
\label{eqn:simple_forw_motion}
    \mathbf{y} = A_{\kappa}\mathbf{x} + \eta, \text{ }\eta \sim \mathcal{N}(0, \sigma^2I), 
\end{equation}
where $A_{\kappa}$ includes all linear operators in Eq. \ref{eqn:motion_forw_less_expanded}. We note here that the motion operators ($R_\theta$,$P_\phi$) are the same for all coils. 

For much of the experimentation in this paper we not only assume constant motion states during a single read out but also fixed motion states for each TR. This is not required for our method to work but it fits with the observation that time between TRs is much longer, in general, than time between readouts within a TR. This is, for example, the case in many fast spin-echo (FSE) imaging sequences. We wish to note that although we explicitly consider rigid body motion in our forward model formulation, non-rigid motion can also be modeled as modifications to the forward model. However, non-rigid motion requires parameterization of a deformation field which greatly increases problem complexity.

\section{Methods}
\label{sec:methods}

\subsection{Accelerated Motion Correction with Diffusion Generative Models}
\label{subsec:ASMC}
As stated above, prior works have shown promising results when using deep generative diffusion models to solve ill-posed inverse problems like sub-sampled image reconstruction \cite{jalal2021robust, martin_score, CHUNG_score}. In most prior work, however, the forward model is assumed to be fixed and known throughout the reconstruction procedure. In our work, however, we assume that our forward operator $A_{\kappa}$ belongs to a restricted class of operators with unknown parameters $\kappa$ which must be learned during inference. Another way of viewing this problem is as posterior sampling from the joint distribution $p(\mathbf{x},\kappa|\mathbf{y})$. Under this joint posterior, we arrive at the following reverse time ODEs:
\begin{equation}
    \label{eqn:posterior_ode_reverse_bayes_x}
    d\bold{x} = \left(\frac{s'(t)}{s(t)}\bold{x} \right. - s^2(t)\sigma'(t)\sigma(t)\left[ \nabla_\bold{x}\log{p(\bold{y}|\Tilde{\bold{x}},\kappa;\sigma(t))} \right.  + \left. \nabla_\bold{x}\log{p(\Tilde{\bold{x}},\kappa;\sigma(t))}\right] \left. \vphantom{\frac{s'(t)}{s(t)}} \right)dt,
\end{equation}
\begin{equation}
    \label{eqn:posterior_ode_reverse_bayes_kappa}
    d\kappa = \left(\frac{s'(t)}{s(t)}\mathbf{\kappa} \right. - s^2(t)\sigma'(t)\sigma(t)\left[ \nabla_\bold{\kappa}\log{p(\bold{y}|\Tilde{\bold{x}},\kappa;\sigma(t))} \right.  + \left. \nabla_\bold{\kappa}\log{p(\Tilde{\bold{x}},\kappa;\sigma(t))}\right] \left. \vphantom{\frac{s'(t)}{s(t)}} \right)dt.
\end{equation}
If we assume independence between $\mathbf x$ and $\kappa$ we arrive at the following reverse ODEs:
\begin{equation}
    \label{eqn:posterior_ode_reverse_bayes_x_ind}
    d\bold{x} = \left(\frac{s'(t)}{s(t)}\bold{x} \right. - s^2(t)\sigma'(t)\sigma(t)\left[ \nabla_\bold{x}\log{p(\bold{y}|\Tilde{\bold{x}},\kappa;\sigma(t))} \right.  + \left. \nabla_\bold{x}\log{p(\Tilde{\bold{x}};\sigma(t))}\right] \left. \vphantom{\frac{s'(t)}{s(t)}} \right)dt,
\end{equation}
\begin{equation}
\label{eqn:posterior_ode_reverse_bayes_kappa_ind}
    d\kappa = \left(\frac{s'(t)}{s(t)}\mathbf{\kappa} \right. - s^2(t)\sigma'(t)\sigma(t)\left[ \nabla_\bold{\kappa}\log{p(\bold{y}|\Tilde{\bold{x}},\kappa;\sigma(t))} \right.  + \left. \nabla_\bold{\kappa}\log{p(\kappa;\sigma(t))}\right] \left. \vphantom{\frac{s'(t)}{s(t)}} \right)dt.
\end{equation}
From here we note that $p(\mathbf{y}|\mathbf{x},\kappa;\sigma(0)) \sim \mathcal{N}(A_\kappa \mathbf{x},\sigma^2 I)$
and we arrive at the final algorithm in Alg. \ref{alg:MI-PS}.

\begin{algorithm}[t]
\caption{General Posterior Sampling With Motion Correction}\label{alg:MI-PS}
\begin{algorithmic}[1]
\Procedure{Gen-MI-PS}{$D_\theta(\bold{x};\sigma(t)),\sigma(t),s(t), t_{i\in{0,...,N}}, \mathbf{y}$}
\State \textbf{sample}
$\mathbf{x}(t_0) \sim \cN(\mathbf{0}, \sigma^2(t_0)s^2(t_0)\mathbf{I})$
\State \textbf{sample}
$\mathbf{\kappa}(t_0) \sim \cN(\mathbf{0}, \sigma^2(t_0)s^2(t_0)\mathbf{I})$
  \For{$i \in \{0,...,N-1\}$}
  \State $\Tilde{\mathbf{x}} = \frac{\mathbf{x}(t_i)}{s(t_i)}$
  
  \State $\nu_i = (\frac{\sigma'(t_i)}{\sigma(t_i)}  + \frac{s'(t_i)}{s(t_i)})$
  \State $\zeta_i = \frac{\sigma'(t_i)s(t_i)}{\sigma(t_i)}$

  \State $\mathbf{d}_{P_x} = \nu_i\mathbf{x}(t_i) -  \zeta_{i} D_\theta\left(\Tilde{\mathbf{x}};\sigma(t_i) \right)$

  \State $\hat{\mathbf{x}} = \frac{1}{s(t_i)}\left(\mathbf{x}(t_i) + s(t_i)^2 \sigma(t_i)^2 \frac{D_\theta\left(\Tilde{\mathbf{x}};\sigma(t_i)\right)-\mathbf{x}(t_i)}{\sigma(t_i)^2}\right)$

  \State $\mathbf{d}_{L_x} = \nabla_{\mathbf{x}} \norm{A_\kappa\hat{\mathbf{x}}(\mathbf{x}(t_i))-\mathbf{y}}_2^2$

  \State $\mathbf{x}(t_{i+1}) \leftarrow \mathbf{x}(t_i) + (t_{i+1}-t_i)\bold{d}_P + \gamma_{t_i} \mathbf{d}_L$

   \State $\mathbf{d}_{P_\kappa} = \nabla_{\kappa} \log{p(\kappa;\sigma(t))}$

  \State $\mathbf{d}_{L_\kappa} = \nabla_{\kappa} \norm{A_\kappa\hat{\mathbf{x}}(\mathbf{x}(t_i))-\mathbf{y}}_2^2$

  \State $\mathbf{\kappa}(t_{i+1}) \leftarrow \kappa(t_i) + (t_{i+1}-t_i)\mathbf{d}_{P_\kappa} + \xi_{t_i} \mathbf{d}_{L_\kappa}$
  \EndFor
\EndProcedure
\end{algorithmic}
\end{algorithm}

\subsection{Comparison Methods}
We display the results for five different methods:
\begin{enumerate}
    \item \textbf{Linear Reconstruction Lower Bound (Linear-LB)}: Conjugate gradient (CG) algorithm-based reconstruction of motion corrupted data with no motion correction (assumes zero motion occurred during scan). We use this as a lower bound for the performance of methods like NAMER \cite{namer}.
    
    \item \textbf{Linear Reconstruction Upper Bound (Linear-UB)}: CG-based reconstruction of motion corrupted data with access to the true motion states. We use this as an upper bound for the performance of methods like NAMER \cite{namer} as in the best case, NAMER is a CG reconstruction with access to ground truth measurements of the motion states.
    \item \textbf{End-to-End Deep Learning (E2E)}: A GAN-based correction method was trained on image pairs of corrupted and clean images. We use a U-Net\cite{unet} and ResNet-18\cite{resnet} model for the generator and the discriminator networks, respectively. To have better visual quality in the generated images, the training loss includes $L1$, adversarial, and  perceptual components through an Imagenet pre-trained VGG model\cite{vgg}.
    During training the generator is given motion corrupted images, from a given sampling pattern and acceleration level, as inputs and trained to generate images as close as possible to the clean images. The network is trained with a learning rate of $0.0001$ for 10 epochs. The input images are normalized prior to passing through the network. 
    \item \textbf{Posterior Sampling (PS)}: Diffusion based posterior sampling reconstruction of motion corrupted data with no motion correction (i.e., assumes zeros motion occurred during scan). The inference procedure for PS is found in Alg. \ref{alg:DPS}
    \item \textbf{Motion Informed Posterior Sampling (MI-PS)}: Diffusion based posterior sampling reconstruction of both image and motion states using Alg. \ref{alg:MI-PS}. 
\end{enumerate}

\subsection{Experiments}
\subsubsection{Simulated Motion}
\begin{figure}[h!]
\begin{center}
  \includegraphics[width=0.70\linewidth]{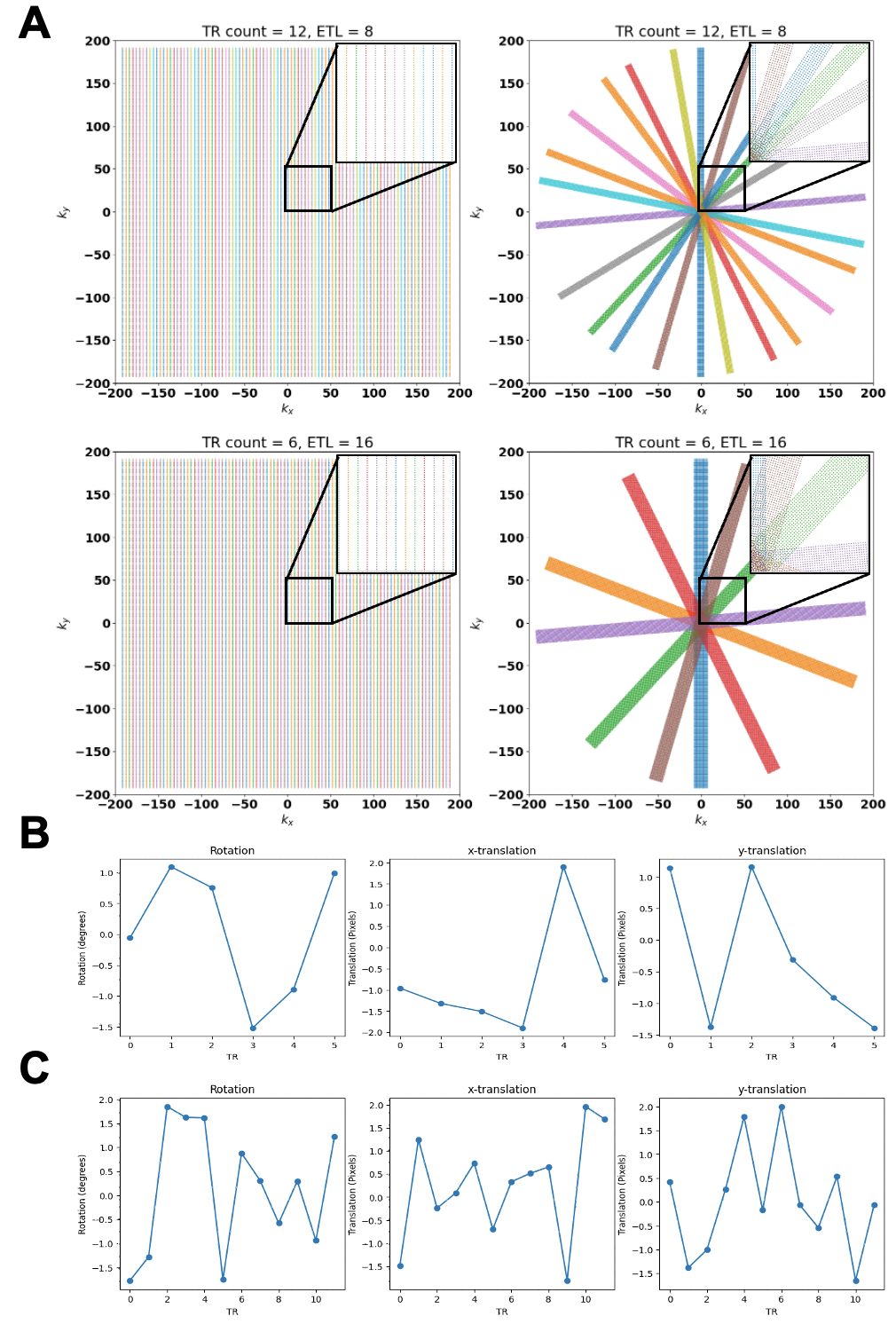}
\end{center}
\caption{(A) For each image shown the differing colors denote separate TRs. For each TR we simulate a single separate motion state triplet (rotation, x-translation, y-translation). (Top Left) $R=4$ Cartesian sampling trajectory $ETL=8$, TR count $=12$. (Bottom Left) $R=4$ Cartesian sampling trajectory $ETL=16$, TR count $=6$.(Top Right) $R=4$ PROPELLER sampling trajectory $ETL=8$, TR count $=12$. (Bottom Right) $R=4$ PROPELLER sampling trajectory $ETL=16$, TR count $=6$. Example motion states. (B) motion trajectory for $R=4$, $ETL=16$ ($\text{TR count} = 6$) sampling pattern. (C) motion trajectory for $R=4$, $ETL=8$ ($\text{TR count} = 12$) sampling pattern.}
\label{fig:trajectories_and_motion}
\end{figure}

To test the robustness of our method at a variety of acceleration levels and sampling patterns, we simulated motion on $100$ T2 brain images from the fastMRI dataset \cite{zbontar2018fastmri} for four different sampling patterns at three different accelerations each. Specifically, we use Cartesian and PROPELLER \cite{propeller} based sampling patterns each at echo train lengths (ETLs) of $8$ and $16$ for accelerations of $R=3,4,5$. All sampling patterns assumed a fully sampled readout of $384$ points. Example trajectories for each sampling pattern and ETL are shown at $R=4$ in Fig.~\ref{fig:trajectories_and_motion}. For each TR we simulate a single independent motion state triplet (rotation, x-translation, y-translation). This means that, for example, the case of Cartesian (or PROPELLER) sampling at $R=4$ with an $ETL=8$ resulted in $12$ TRs and thus $12$ motion states to estimate along with the corrected image. The motion states for each TR were sampled independently from a uniform distribution ($\kappa \sim \mathcal{U}(-2,2)$). See Fig. \ref{fig:trajectories_and_motion} for an example of simulated motion states for a given $(R,ETL)$ pair.

Prior to simulating motion corruption over the raw k-space data we first resized the fully sampled k-space to be $384\times 384$. Next we calculated sensitivity maps using ESPIRiT \cite{ESPIRiT}. We then applied motion to k-space measurements by drawing random motion states and passing the coil images through the motion-corrupt forward operator. Finally, we added a small amount of noise to the sampled k-space data. We note here that many parts of the preprocessing here constitute an inverse crime \cite{efrat_pnas}. However, all competing methods used the same data so our method should not have gained an unfair advantage in this respect. 

\subsubsection{Prospective Motion}
As proof of principle on prospective data we acquired a single T2 brain scan on a healthy volunteer with institutional review board approval and informed consent. The data were collected on a Siemens Vida 3 Tesla MRI scanner at our institution, and we emphasize that the scanner hardware and imaging protocol differed from the fastMRI training data. We first collected a scan at $R=2$ where we asked the participant to not move during the scan and consider this our motion free scan. Finally, we collected a scan at $R=3$ where we asked the participant to rotate and translate their head (approximately) in-plane during the scan. Scan parameters were: ETL=$16$, R=$3$, slice thickness = $4mm$, FOV=$220mm \times 220mm$, resolution=$0.42mm\times0.42mm$ .We applied the LINEAR-LB, PS, and MI-PS methods to compare their reconstructions on prospective data. For MI-PS, instead of estimating one motion state for each TR we estimated separate motion states for each phase encode.

\subsection{Implementation Details}\label{sec:implementation}

\subsubsection{Training}
To train our deep diffusion network ($D_\theta(\mathbf{x},\sigma(t))$) we used $14,000$ samples from the fastMRI brain dataset \cite{zbontar2018fastmri}. The model was a UNet style architecture with $2$ input/output channels for real and imaginary components of the images. The network contained $64,675,076$ trainable parameters. We used a batch size of $15$ for training the diffusion model and training was done across three $A100$ GPUs. The E2E method was trained using simulated motion corrupt samples from a Cartesian trajectory with $R=3$, $ETL=16$. Training for this model was done on a single $A40$ GPU. 

\subsubsection{Inference}
We selected the following parameters to define the drift and diffusion coefficients in the forward process: $s(t) = 1$, $\sigma(t)=t$ with $\sigma_{max} = 5$ and $\sigma_{min}=0.002$. For running the reverse ODE we also selected a total number of inference time steps $N=300$ and a time step schedule of $t_i = \left(\sigma_{max}^{1/\rho} + \frac{i}{N-1}\left(\sigma_{min}^{\frac{1}{\rho}}-\sigma_{max}^{\frac{1}{\rho}}\right) \right)^\rho$ where $\rho=7$. Similar to \cite{chung2023dps} we select the likelihood weighting $\gamma_{t_i} = \frac{1}{\norm{A_\kappa\hat{x}-y}}_2$. As motion was simulated, we did not wish to unfairly assume a prior over motion states. Therefore we set $\mathbf{d}_{P_\kappa} = 0$. Finally we used a fixed step size ($\xi$) for updating motion estimates. For Cartesian sampling patterns $\xi = 1$ while for PROPELLER $\xi=0.3$ was used. Finally, we found it best to initialize the motion estimates to zero ($\kappa = 0$). These details lead to the final update procedure shown in Alg. \ref{alg:final_alg}.

\begin{algorithm}[t]
\caption{Posterior Sampling With Motion Correction}\label{alg:final_alg}
\begin{algorithmic}[1]
\Procedure{MI-PS}{$D_\theta(\bold{x};\sigma(t)),\sigma(t),s(t), t_{i\in{0,...,N}}, \mathbf{y}$}
\State \textbf{sample}
$\mathbf{x}(t_0) \sim \cN(\mathbf{0}, \sigma^2(t_0)s^2(t_0)\mathbf{I})$
\State $\mathbf{\kappa}(t_0) = 0$
  \For{$i \in \{0,...,N-1\}$}

  \State $\hat{\mathbf{x}} =   D_\theta\left(\mathbf{x};\sigma(t_i)\right)$
  
  \State $\mathbf{d}_{P_x} = \frac{\mathbf{x}(t_i) -   \hat{\mathbf{x}}}{t_i} $

  \State $\mathbf{d}_{L_x} = \nabla_{\mathbf{x}} \norm{A_\kappa\hat{\mathbf{x}}(\mathbf{x}(t_i))-\mathbf{y}}_2^2$

  \State $\mathbf{d}_{L_\kappa} = \nabla_{\kappa} \norm{A_\kappa\hat{\mathbf{x}}(\mathbf{x}(t_i))-\mathbf{y}}_2^2$

  \State $\mathbf{x}(t_{i+1}) \leftarrow \mathbf{x}(t_i) + (t_{i+1}-t_i)\bold{d}_P + \gamma_{t_i} \mathbf{d}_L$

  \State $\mathbf{\kappa}(t_{i+1}) \leftarrow \kappa(t_i) + \xi_{t_i} \mathbf{d}_{L_\kappa}$
  \EndFor
\EndProcedure
\end{algorithmic}
\end{algorithm}

\subsection{Quantitative Evaluation}
We evaluate the retrospective results using normalized root mean squared error (NRMSE) and structural similarity index measure (SSIM) on a validation set of 100 images. As there is an ambiguity between two data-consistent reconstructions with a fixed motion offset, we first align the reconstruction to the motion-free ground-truth before evaluating NRMSE and SSIM.

\section{Results}
\begin{figure*}[h!]
  \begin{center}
      \includegraphics[width=1.\linewidth]{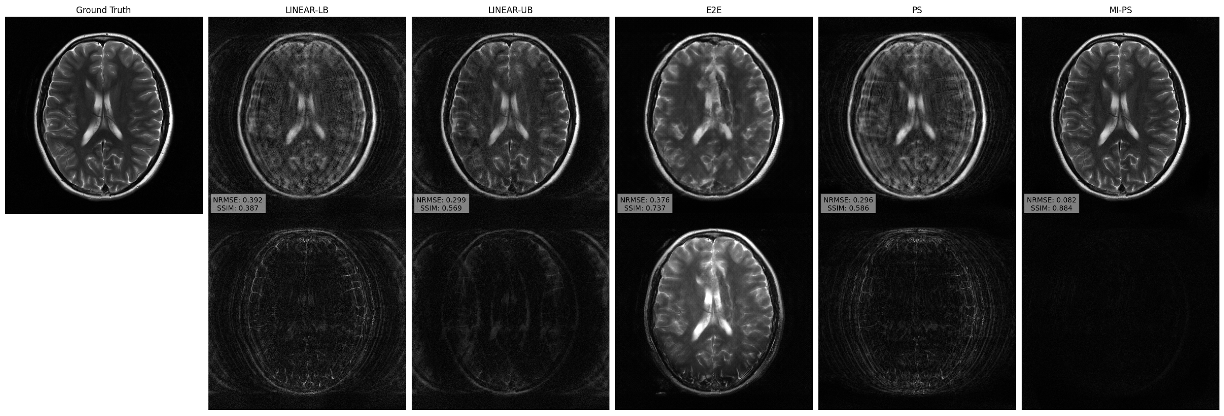}
  \end{center}
\caption{Example reconstruction of simulated motion for a Cartesian sampling pattern with $R=4$, $ETL = 16$. From left to right: motion-free fully sampled ground-truth; LINEAR-UB; LINEAR-LB; end-to-end GAN; posterior sampling without motion correction, proposed motion-informed posterior sampling.}
\label{fig:sim_cart_recon}
\end{figure*}

\begin{figure*}[h!]
  \begin{center}
      \includegraphics[width=1.\linewidth]{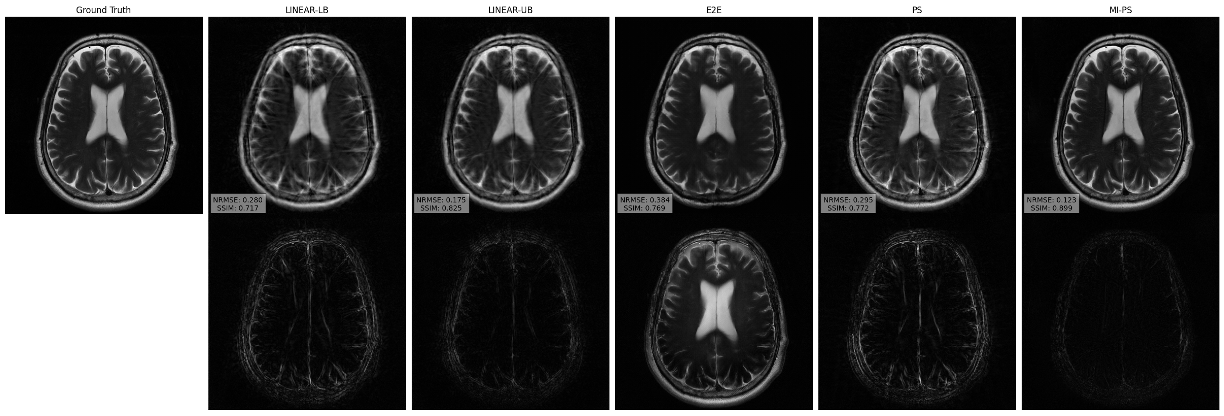}
  \end{center}
\caption{Example reconstruction of simulated motion for a PROPELLER sampling pattern with $R=5$, $ETL = 8$. From left to right: motion-free fully sampled ground-truth; LINEAR-UB; LINEAR-LB; end-to-end GAN; posterior sampling without motion correction, proposed motion-informed posterior sampling. }
\label{fig:sim_prop_recon}
\end{figure*}

Quantitative NRMSE and SSIM metrics are shown for each simulated motion case in Table \ref{table:sim_results}.  Example reconstructions for simulated motion using Cartesian and PROPELLER trajectories can be seen in in Fig. \ref{fig:sim_cart_recon} and \ref{fig:sim_prop_recon} respectively. Reconstruction without accounting for motion leads to large error, necessitating the use of motion correction. However, the LINEAR-UB reconstruction, which has access to ground-truth motion parameters performs poorly due to residual aliasing artifacts. While the E2E method is able to remove aliasing, residual motion artifacts remain, likely because of the differences in motion at test time. Posterior sampling without motion correction is not able to compensate for motion. Finally, MI-PS is able to handle the acceleration as well as the motion through joint posterior sampling.   

Using a PROPELLER based acquisition leads to lower error for LIENAR-UB and LINEAR-LB, likely due to the presentation of incoherent artifacts from subsampling and from motion. Interestingly, PROPELLER reconstructions are somewhat worse quantitatively for posterior sampling compared to Cartesian acquisition. This may be due to the heavier subsampling of high frequency k-space.

\begin{figure*}[h!]
  \begin{center}
      \includegraphics[width=1.\linewidth]{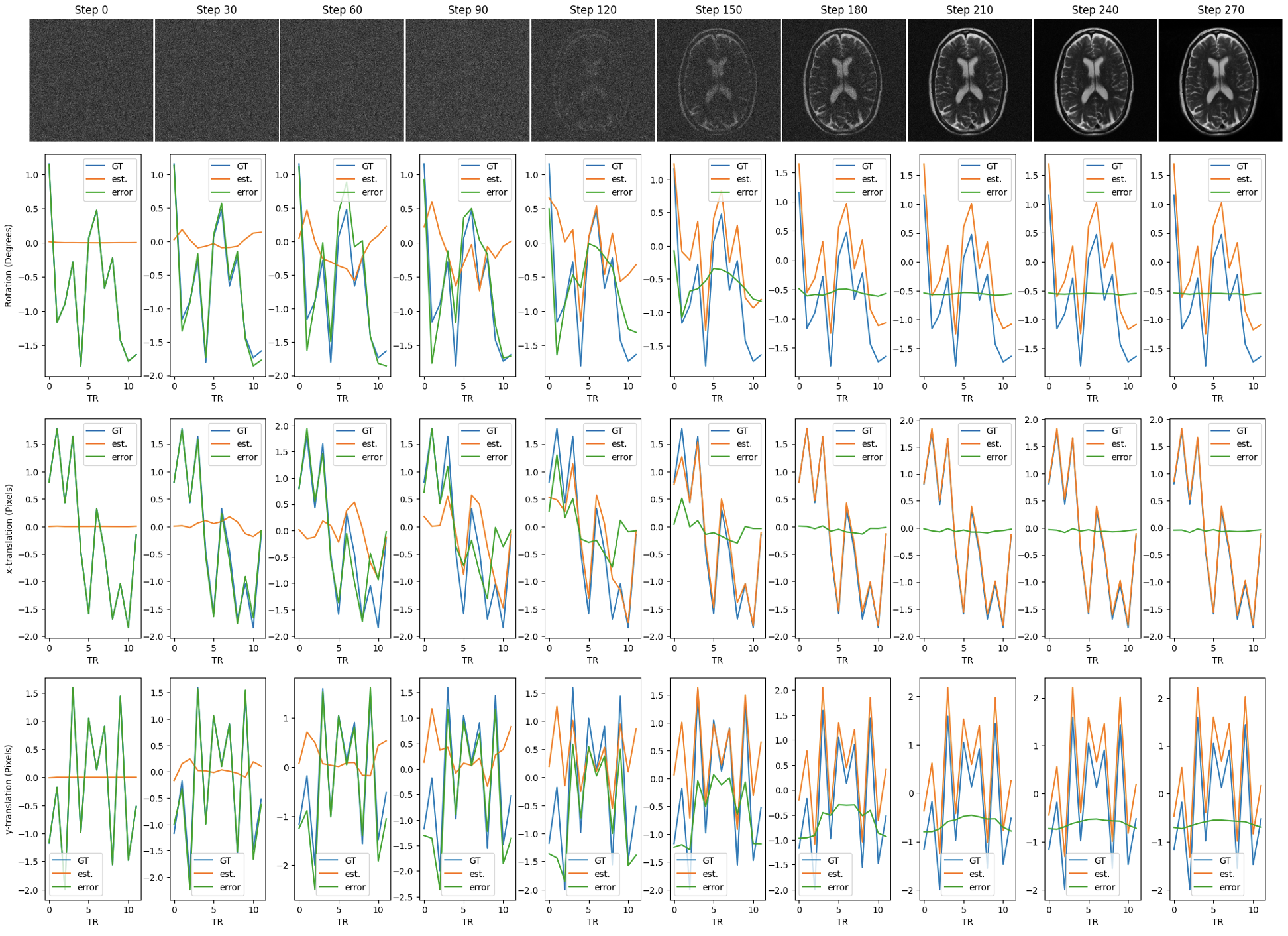}
  \end{center}
\caption{Progression of image ($\mathbf{x}$) and motion ($\kappa$) estimates during inference. Iterations are shown from left to right. Motion estimates and error curves are shown for 2D rigid motion parameters. }
\label{fig:prog}
\end{figure*}

In Fig. \ref{fig:prog} we show the stochastic dynamics of MI-PS over the reconstruction iterations. The image is initialized to all-noise and the motion states are initialized to zero. As the iterations progress, the image and motion jointly converge to a stable solution. The motion states show close alignment with the ground-truth motion states except for a fixed offset as previously discussed.

\begin{figure*}[h!]
  \begin{center}
  \includegraphics[width=1.0\linewidth]{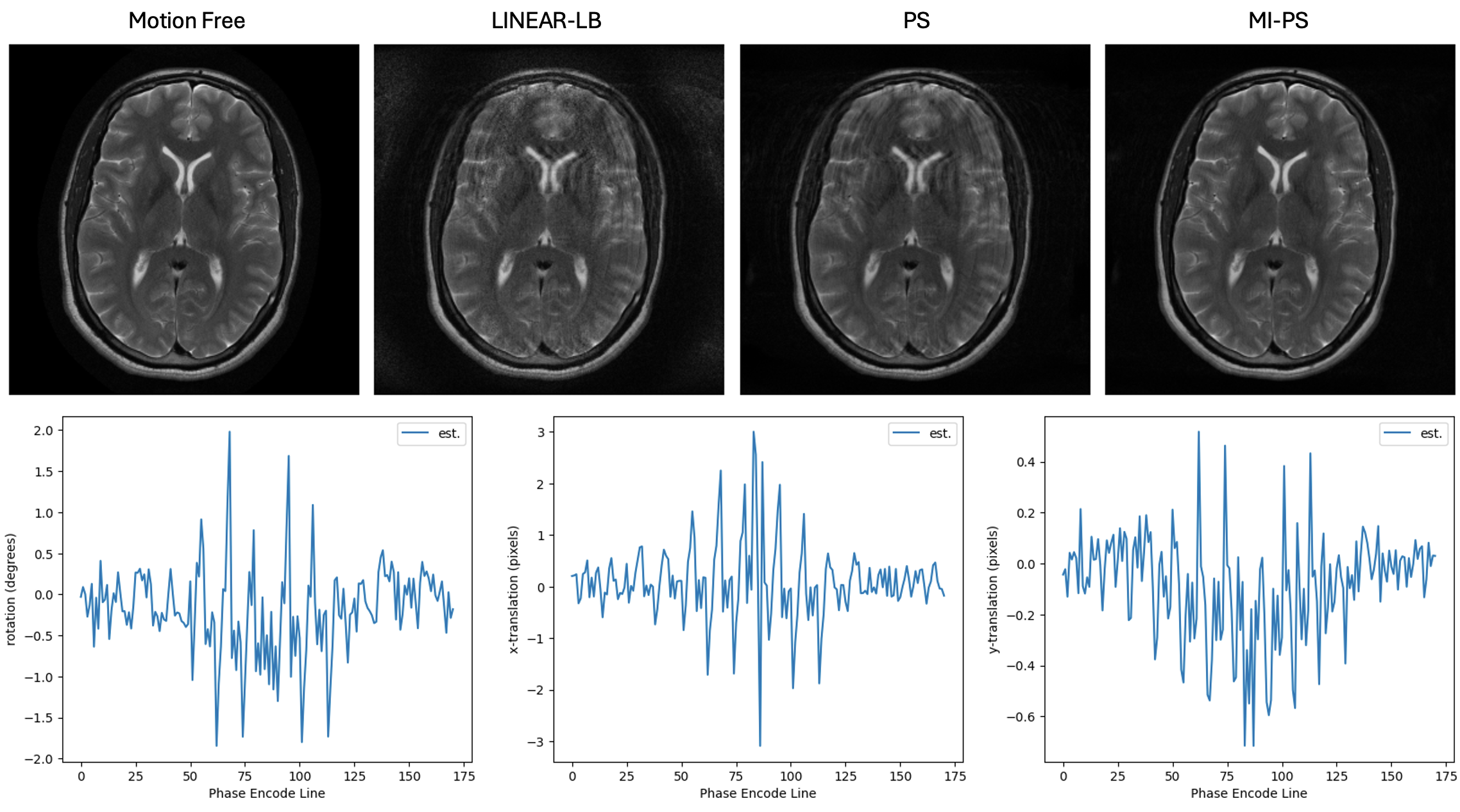}
  \end{center}
\caption{(Top) Reconstruction results for prospectively accelerated scan at $R=3$ for a subset of the tested methods. The motion-free scan was performed at R=2 with linear reconstruction. LINEAR-LB, posterior sampling, and MI-PS were reconstructed from motion-corrupt scan at R=3.  (Bottom) motion state estimates for each phase encode line using our method.}
\label{fig:prosp_recon_ex}
\end{figure*}

Finally, we demonstrate a subset of the  reconstruction techniques on the prospectively accelerated scan in Fig. \ref{fig:prosp_recon_ex}. We display the final image reconstructions alongside the estimated rigid motion parameters from MI-PS. The motion-free scan was collected independently at a low acceleration and serves as a qualitative baseline. LINEAR-LB demonstrates parallel imaging reconstruction alone, indicating the impact of the motion during the scan. Posterior sampling alone is not able to remove these motion artifacts and thus qualitatively matches LINEAR-LB. Finally, MI-PS is able to remove the motion artifacts by estimating the motion states. Notably, the motion estimates for outer k-space are lower in magnitude compared to the low-frequency k-space points, likely due to poorer estimation due to lower signal-to-noise (SNR) ratio.

\begin{table*}[t]
\footnotesize
\centering
\caption{Simulated Motion Results}
\label{table:sim_results}
\begin{tabularx}{1\linewidth}{ c *{11}{c} }
\\
    \toprule
   \textbf{R} & \multicolumn{2}{c}{\textbf{LINEAR-LB}}
                            & \multicolumn{2}{c}{\textbf{LINEAR-UB}}
                            & \multicolumn{2}{c}{\textbf{E2E}}
                            & \multicolumn{2}{c}{\textbf{PS}}
                            & \multicolumn{2}{c}{\textbf{MI-PS}}\\
   &   NRMSE$\downarrow$  & 
   SSIM$\uparrow$  &
   NRMSE$\downarrow$  &
   SSIM$\uparrow$ &
   NRMSE$\downarrow$ &
   SSIM$\uparrow$&
   NRMSE$\downarrow$ &
   SSIM$\uparrow$&
   NRMSE$\downarrow$ &
   SSIM$\uparrow$\\
    \midrule
    \multicolumn{11}{c}{\textbf{Cartesian}, $ETL = 8$}
    \\
    \midrule
$3$   &   $0.348$   &   $0.423$ & $0.204$  &   $0.646$ &  $0.271$ &  $0.805$  & $0.293$ & $0.546$ & $\mathbf{0.072}$ & $\mathbf{0.892}$
\\
$4$   &   $0.401$   &   $0.346$ & $0.305$  &   $0.561$ &  $0.374$ &  $0.725$  & $0.299$ & $0.558$ & $\mathbf{0.088}$ & $\mathbf{0.871}$
\\
$5$   &   $0.421$   &   $0.354$ & $0.338$  &   $0.509$ &  $0.405$ &  $0.687$  & $0.298$ & $0.562$ & $\mathbf{0.094}$ & $\mathbf{0.859}$\\
\midrule
    \multicolumn{11}{c}{\textbf{Cartesian}, $ETL = 16$}
\\
\midrule
$3$   &   $0.341$   &   $0.442$ & $0.204$  &   $0.642$ &  $0.279$ &  $0.802$  & $0.293$ & $0.546$ & $\mathbf{0.072}$ & $\mathbf{0.897}$
\\
$4$   &   $0.396$   &   $0.385$ & $0.304$  &   $0.561$ &  $0.376$ &  $0.725$  & $0.298$ & $0.772$ & $\mathbf{0.085}$ & $\mathbf{0.881}$
\\
$5$   &   $0.434$   &   $0.359$ & $0.364$  &   $0.501$ &  $0.447$ &  $0.659$  & $0.291$ & $0.582$ & $\mathbf{0.099}$ & $\mathbf{0.868}$\\
\midrule
\multicolumn{11}{c}{\textbf{PROPELLER}, $ETL = 8$}
\\
\midrule
$3$   &   $0.287$   &   $0.721$ & $0.123$  &   $\mathbf{0.889}$ &  $0.358$ &  $0.766$  & $0.301$ & $0.701$ & $\mathbf{0.111}$ & $0.858$
\\
$4$   &   $0.293$   &   $0.715$ & $0.150$  &   $0.861$ &  $0.368$ &  $0.758$  & $0.307$ & $0.713$ & $\mathbf{0.116}$ & $\mathbf{0.871}$
\\
$5$   &   $0.305$   &   $0.703$ & $0.183$  &   $0.823$ &  $0.378$ &  $0.748$  & $0.315$ & $0.725$ & $\mathbf{0.127}$ & $\mathbf{0.878}$\\
\midrule
\multicolumn{11}{c}{\textbf{PROPELLER}, $ETL = 16$}
\\
\midrule
$3$   &   $0.285$   &   $0.745$ & $0.124$  &   $\mathbf{0.893}$ &  $0.368$ &  $0.761$  & $0.297$ & $0.767$ & $\mathbf{0.101}$ & $0.885$
\\
$4$   &   $0.288$   &   $0.744$ & $0.158$  &   $0.863$ &  $0.374$ &  $0.752$  & $0.298$ & $0.772$ & $\mathbf{0.114}$ & $\mathbf{0.885}$
\\
$5$   &   $0.304$   &   $0.720$ & $0.225$  &   $0.802$ &  $0.391$ &  $0.739$  & $0.299$ & $0.770$ & $\mathbf{0.161}$ & $\mathbf{0.868}$\\

    \bottomrule
\end{tabularx}
\end{table*}

\section{Discussion}
\label{sec:discussion}
Deep diffusion generative models have shown to be an extremely powerful advancement for solving inverse problems due to their ability to decouple the measurement model from the prior, which can be well-modeled with deep neural networks. In particular, this means that in contrast to end-to-end inverse problem solvers, diffusion priors can be used in a more modular fashion for a variety of imaging problems which vary based on their measurement forward operator (e.g., accelerated reconstruction, super resolution, etc.). In many inverse problem settings, however, the true form of the forward operator used to collect measurements may be unknown as is the case in motion corrupt MRI scans.  Several prior methods have approached the problem of motion correction by incorporating the parameterization of motion in the forward operator and jointly optimizing over both the image and motion variables \cite{sense_encode,tamer,namer}.

In a similar fashion, our proposed approach solves this problem by treating the unknown parameterization of the motion as a complementary random variable which must be jointly sampled alongside the clean MR image of interest from motion corrupted measurements (i.e., $\mathbf{x},\kappa \sim p(\mathbf{x},\kappa | y)$). This is similar to other recently proposed techniques for solving blind inverse problems \cite{Chung_2023_CVPR,murata2023gibbsddrm}. This treatment of the motion correction problem enables us to decouple the training of the prior from the motion correction task which allows our method to be transferable between differing sampling trajectories. This is extremely important as the manifestation of motion artifacts in the final image is highly dependent on the sampling trajectory used to collect measurements \cite{FSE_comp_moco}. It is this property of motion artifacts that makes it difficult to generalize end-to-end methods to arbitrary motion corruption.

We see that our method outperforms both end-to-end methods and the best-case linear reconstructions for prior joint optimization techniques \cite{tamer,namer}. This can be owed to the powerful image prior provided by the diffusion generative model which discourages motion estimates that give rise to motion corrupt images at reconstruction time. Although we did not use a trained prior to regularize our motion estimates we implicitly used the prior of enforcing all motion states in the same TR to be the same. We showed on prospectively acquired data that we can drop this assumption and still obtain good reconstruction results. However, it is clear from Fig. \ref{fig:prosp_recon_ex} that motion estimates for outer k-space differ from those in low-frequency k-space, despite the T2-weighted echo train ordering. This indicates that a more informative prior, e.g. by explicitly grouping the phase encodes in a single ETL, may improve reconstruction quality.

Recently, adaptive end-to-end methods have been proposed under the lens of hybrid networks. In these settings, motion parameters are estimated and used to choose the weights of the neural network that was trained for those weights \cite{nalini_motion}. Similarly, unrolled methods that explicitly solve for motion using principled optimization have also been proposed \cite{hammernik_motion,kamilov_motion}. These methods may provide a solution to out-of-distribution error, though they still must be re-trained for different sampling trajectories.

We assumed a flat prior on the motion states, which is quite naive. For example, we do not account for the likely causal transition between one motion state and the next. A stronger prior on the motion, for example through a Gaussian process or other Markov chain, could be a flexible way to model such time-dependencies. However, this may not be suitable for rapid and sporadic motion. For these reason, we chose to keep the prior simple. The effects of the design choice should be explored further to determine the potential benefits of richer motion priors.

Our prospective scan result sheds light on the power of our framework when the training set does not match the test set -- both in protocol settings as well as scanner hardware. In essence, it is not necessary to retrain the model for every specific MRI protocol. However, we note that we only scanned a single slice, whereas nearly all MRI protocols will collect multi-slice data. While in principle this could be handled by our framework, we did not apply it to multi-slice data. Our framework was also formulated for 2D rigid motion only. Therefore, we are not able to model out-of-plane motion or nonrigid motion. Our framework could be extended to 3D rigid motion estimation by adding additional rotation and translation parameters for each time step. Nonrigid motion could also be handled through proper parameterization of the nonrigid space, e.g. thorough spline interpolation or optical flow.

Posterior sampling is a promising alternative to image reconstruction from subsampled measurements compared to other point estimators. One reason is for the ability to quantify uncertainty in the result. Similar to conventional posterior sampling, joint posterior sampling over image and motion cold also be used to learn uncertainty in the motion parameters. In this work we did not explicitly explore uncertainty in the motion, though it could be interesting for future work.


\section{Conclusion}
\label{sec:conclusion}

We proposed a method to correct motion artifacts from highly accelerated MRI by parameterizing motion as inconsistencies in the forward operator which can be jointly estimated as random variables alongside a clean reconstructed MR image. To solve the joint estimation problem we leveraged advancements in deep generative diffusion models to perform posterior sampling. We displayed our proposed technique's ability to correct 2D rigid body motion on both simulated and prospectively corrupted scan data.

\section{Acknowledgment}
This work was supported by Aspect Imaging, Google Research Scholar Program, NSF IFML 2019844, NIH U24EB029240, NSF CCF-2239687
(CAREER), Oracle for Research Fellowship, and ARO W911NF2110117.

\section{Data Availability Statement}
In the spirit of reproducible research, our source code can be found at \url{https://github.com/utcsilab/motion_score_mri}.

\section{Conflicts of Interest}
The authors do not have any conflicts of interest.

\section{ORCID}
\textit{Brett Levac} \url{https://orcid.org/0000-0002-1182-3438}\\
\noindent\textit{Jonathan I.\ Tamir} \url{https://orcid.org/0000-0001-9113-9566}
\noindent\textit{Ajil Jalal} \url{https://orcid.org/0009-0006-9244-8575}

\clearpage

\clearpage

\end{document}